\documentclass[sigconf,nonacm]{acmart}
\usepackage{listings}
\usepackage{multicol}

\usepackage{listings, xcolor}

\definecolor{verylightgray}{rgb}{.97,.97,.97}

\lstdefinelanguage{Solidity}{
	keywords=[1]{anonymous, assembly, assert, balance, break, call, callcode, case, catch, class, constant, continue, constructor, contract, debugger, default, delegatecall, delete, do, else, emit, event, experimental, export, external, false, finally, for, function, gas, if, implements, import, in, indexed, instanceof, interface, internal, is, length, library, log0, log1, log2, log3, log4, memory, modifier, new, payable, pragma, private, protected, public, pure, push, require, return, returns, revert, selfdestruct, send, solidity, storage, struct, suicide, super, switch, then, this, throw, transfer, true, try, typeof, using, value, view, while, with, addmod, ecrecover, keccak256, mulmod, ripemd160, sha256, sha3}, 
	keywordstyle=[1]\color{blue}\bfseries,
	keywords=[2]{address, bool, byte, bytes, bytes1, bytes2, bytes3, bytes4, bytes5, bytes6, bytes7, bytes8, bytes9, bytes10, bytes11, bytes12, bytes13, bytes14, bytes15, bytes16, bytes17, bytes18, bytes19, bytes20, bytes21, bytes22, bytes23, bytes24, bytes25, bytes26, bytes27, bytes28, bytes29, bytes30, bytes31, bytes32, enum, int, int8, int16, int24, int32, int40, int48, int56, int64, int72, int80, int88, int96, int104, int112, int120, int128, int136, int144, int152, int160, int168, int176, int184, int192, int200, int208, int216, int224, int232, int240, int248, int256, mapping, string, uint, uint8, uint16, uint24, uint32, uint40, uint48, uint56, uint64, uint72, uint80, uint88, uint96, uint104, uint112, uint120, uint128, uint136, uint144, uint152, uint160, uint168, uint176, uint184, uint192, uint200, uint208, uint216, uint224, uint232, uint240, uint248, uint256, var, void, ether, finney, szabo, wei, days, hours, minutes, seconds, weeks, years},	
	keywordstyle=[2]\color{teal}\bfseries,
	keywords=[3]{block, blockhash, coinbase, difficulty, number, timestamp, msg, gas, sender, sig, value, now, tx, gasprice, origin},	
	keywordstyle=[3]\color{violet}\bfseries,
	identifierstyle=\color{black},
	sensitive=false,
	comment=[l]{//},
	morecomment=[s]{/*}{*/},
	commentstyle=\color{gray}\ttfamily,
	stringstyle=\color{red}\ttfamily,
	morestring=[b]',
	morestring=[b]"
}

\usepackage{listings, xcolor}

\definecolor{verylightgray}{rgb}{.97,.97,.97}

\lstdefinelanguage{KeDSL}{
	keywords=[1]{statusCode,callData,output,requires,contract_storage,refund,nonce_slot,is_owner_mapping_slot,txtype_hash,threshold,call_pc,call_log_pcset,domain_separator_slot,readLog,writeLog,callLog,ensures}, 
	keywordstyle=[1]\color{blue}\bfseries,
	keywords=[2]{andBool,notBool,orBool},	
	keywordstyle=[2]\color{teal}\bfseries,
	keywords=[3]{\[fn-execute\],\[fn-execute-a0gt0\],\[fn-execute-a0le0\],\[fn-execute-sig1-invalid\],\[fn-execute-sigs-valid\],\[fn-execute-sig0-invalid\],\[fn-execute-success\],\[fn-execute-failure\],\[fn-execute-overflow\],\[fn-execute-nooverflow\],\[fn-execute-a0lt10\],\[execute\],\[execute-executor-invalid\],\[execute-executor-valid\],\[execute-executor-valid-sigcheck-fail-revert-0\],\[execute-executor-valid-sigcheck-fail-revert-1\],\[execute-executor-valid-sigcheck-fail-revert-2\],\[execute-executor-valid-sigcheck-pass\],\[execute-executor-valid-sigcheck-pass-ownercheck-fail-revert\],\[execute-executor-valid-sigcheck-pass-ownercheck-pass-call\],\[execute-executor-valid-sigcheck-pass-ownercheck-pass-call-success\],\[execute-executor-valid-sigcheck-pass-ownercheck-pass-call-failure\],\[trusted\]},	
	keywordstyle=[3]\color{violet}\bfseries,
    otherkeywords = {;,<<,>>,=>,==Int,+Int,^Int,-Int,>=Int,=/=Int,<Int},
    morekeywords = [2]{=>,_,==Int,+Int,^Int,-Int,>=Int,=/=Int,<Int},
    alsoletter={\#,-,[,]},
	identifierstyle=\color{black},
	sensitive=false,
	comment=[l]{//},
	morecomment=[s]{/*}{*/},
	commentstyle=\color{gray}\ttfamily,
	stringstyle=\color{red}\ttfamily,
	morestring=[b]',
	morestring=[b]"
}

\usepackage{listings, xcolor}

\definecolor{verylightgray}{rgb}{.97,.97,.97}

\lstdefinelanguage{KYaml}{
	keywords=[1]{macro, spec,state,if,then,where,rule,inherits, name, -, match, not, or},	
	keywordstyle=[1]\color{blue}\bfseries,
	keywords=[2]{statusCode,callData,output,storage,refund,readLog,writeLog,callLog}, 
	keywordstyle=[2]\color{teal}\bfseries,
	keywords=[3]{\[trusted\]},	
	keywordstyle=[3]\color{violet}\bfseries,
    otherkeywords = {;,<<,>>,=>,==Int,+Int,^Int,-Int,>=Int,=/=Int,<Int,==K},
    morekeywords = [2]{=>,_,==Int,+Int,^Int,-Int,>=Int,=/=Int,<Int,-,==K},
    alsoletter={\#,-,[,]},
	identifierstyle=\color{black},
	sensitive=false,
	comment=[l]{//},
	morecomment=[s]{/*}{*/},
	commentstyle=\color{gray}\ttfamily,
	stringstyle=\color{red}\ttfamily,
	morestring=[b]',
	morestring=[b]"
}

\usepackage{listings, xcolor}

\definecolor{verylightgray}{rgb}{.97,.97,.97}

\lstdefinelanguage{K}{
	keywords=[1]{syntax,rule,requires}, 
	keywordstyle=[1]\color{blue}\bfseries,
	keywords=[2]{andBool},	
	keywordstyle=[2]\color{teal}\bfseries,
	keywords=[3]{\[function\],,\[trusted\]},	
	keywordstyle=[3]\color{violet}\bfseries,
    otherkeywords = {;,<<,>>,+,=>,::=},
    morekeywords = [2]{=>,+,_},
    morekeywords = [3]{::=},
    alsoletter={\#,-,[,]},
	identifierstyle=\color{black},
	sensitive=false,
	comment=[l]{//},
	morecomment=[s]{/*}{*/},
	commentstyle=\color{gray}\ttfamily,
	stringstyle=\color{red}\ttfamily,
	morestring=[b]',
	morestring=[b]"
}

\AtBeginDocument{%
  \providecommand\BibTeX{{%
    \normalfont B\kern-0.5em{\scshape i\kern-0.25em b}\kern-0.8em\TeX}}}

\begin{document}

\title{User Experience with Language-Independent Formal Verification}

\author{Suhabe Bugrara}
\affiliation{%
  \institution{ConsenSys}
}

\begin{abstract}
The goal of this paper is to help mainstream programmers routinely use formal verification on their smart contracts by 1) proposing a new YAML-format for writing general-purpose formal specifications, 2) demonstrating how a formal specification can be incrementally built up without needing advanced training, and 3) showing how formal specifications can be tested by using program mutation.
\end{abstract}

\maketitle

\section{Introduction}
The goal of this paper is to help mainstream programmers routinely use formal verification on their smart contracts. It attempts to achieve this by:
\begin{enumerate}
    \item  proposing a simple, general-purpose, easy-to-read YAML format for writing formal specifications that is purely syntactic sugar for the K Framework~\cite{Rosu:2010} and can thus, in principle, be used for any smart contract language (Section~\ref{sec:kyaml}), 
    \item demonstrating how a programmer can incrementally build up a formal specification for their program without advanced training and without needing to understand the internals of the formal verification prover (Sections~\ref{sec:walkthrough}),
    \item showing how formal specifications can be tested by constructing \emph{mutant} variations of the program. (Section~\ref{sec:testing}).
\end{enumerate}

Section~\ref{sec:kframework} gives an overview of the  K Framework. Section~\ref{sec:usability} discusses the major usability challenges that state-of-the-art formal verification systems share that have inhibited their mainstream adoption. Section~\ref{sec:simplemultisig} describes the partial formal specification written for the \verb|SimpleMultiSig|, a real Ethereum multisig smart contract. 

\section{K Framework}
\label{sec:kframework}

The K Framework~\cite{Rosu:2010} is a disruptive system in programming languages that seeks to enable the design and implementation of programming tools (such as compilers, virtual machines, deductive verifiers, and others) independently of a particular programming language. Its vision is that, once the syntax and operational semantics of a programming language is specified in K, the K Framework can automatically generate programming tools in a correct-by-construction manner:

\includegraphics[width=.9\linewidth]{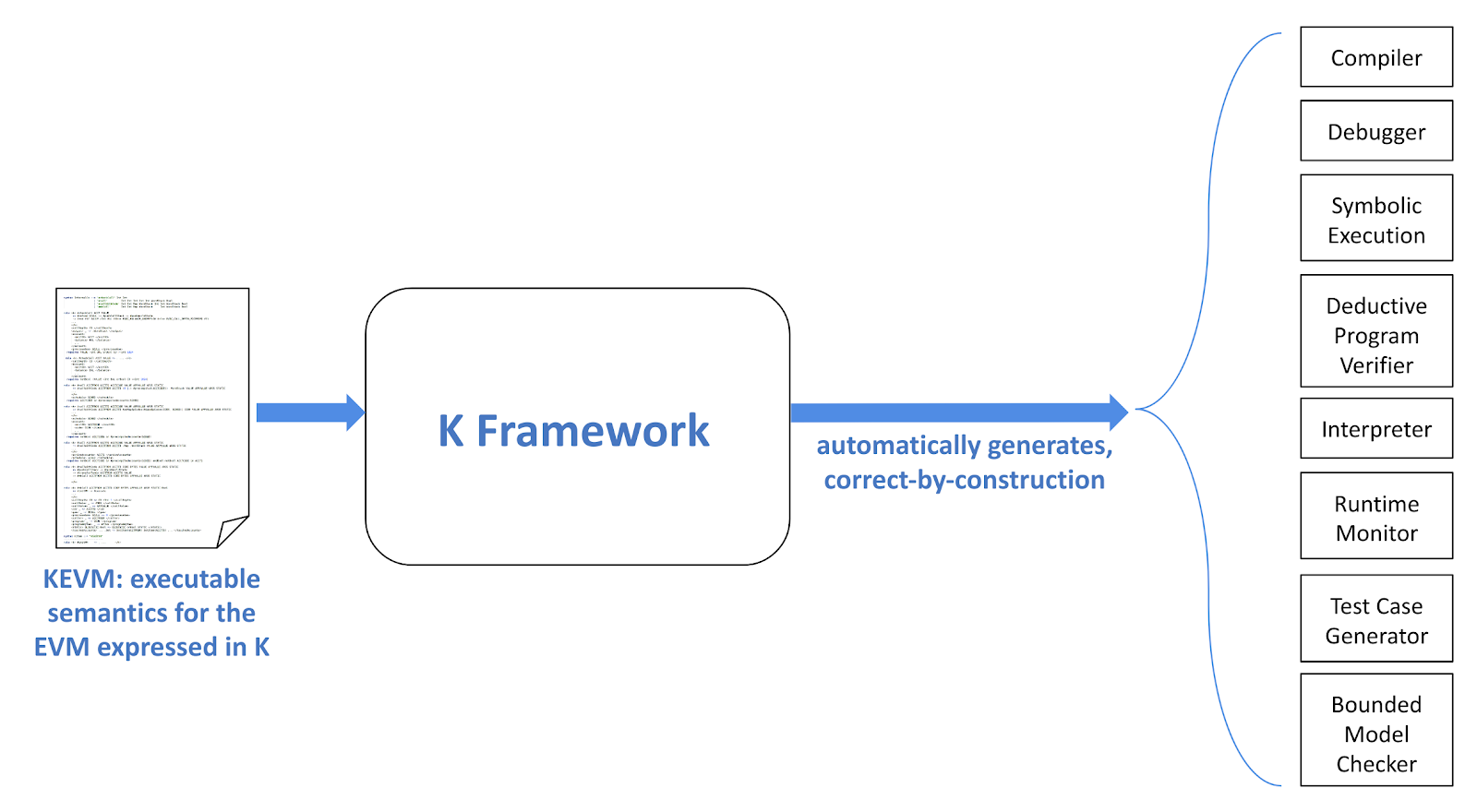}

K semantics have been successfully written and tested for several popular and widely-used languages including C~\cite{Hathhorn:2015}, Java~\cite{Bogdanas:2015}, EVM~\cite{Hildenbrandt:2018}, Javascript~\cite{Park:2015}, x86-64~\cite{Dasgupta:2019}, and others. The K Framework itself and these individual language semantics are open-sourced under the UIUC License, which is permissive and free. One of the distinctive advantages of a K semantics for a programming language is that it is \emph{executable} in the sense that the K Framework can immediately generate an interpreter for the language directly from the K semantics. This interpreter can be used to execute any program written in that language which enables one to actually write and run test cases on the semantics. In contrast, for example, the EVM Yellow Paper~\cite{Wood:YellowPaper} is an English language semantics that is not executable and has been found to be unclear, under-specified, and, in exceptional cases, inconsistent with actual EVM implementations~\cite{Hildenbrandt:2018}. The Ethereum Foundation has been considering adopting the K-EVM semantics as the official semantics for the EVM. 

Even at this early stage of the area, many new programming languages have already been proposed, developed or adopted to support decentralization. Some examples are Move, Mokoto, Solidity, Vyper, Serpent, and WebAssembly. It is impractical and wasteful to repeatedly build a new formal verification system for each language. Furthermore, programming languages are constantly evolving and continually needing to keep a formal verification system up-to-date with the latest versions is expensive. The power of the K Framework is that formal verification systems can be developed \emph{independently} of a particular programming language. Once the semantics of the language is formally defined in K, a K-based prover can immediately be used to formally verify properties of programs written in that language.

 

\section{Usability}
\label{sec:usability}
This section discusses some of the usability issues that state-of-the-art formal verification system typically have. 


\subsection{User Interactions}
\label{sec:user-interactions}
A user typically interacts with a formal verification system in several different ways:

\begin{itemize}
\item The user writes a \emph{formal specification} of the program in a declarative, logic-based specification language that the prover must be designed to handle.
\item The user writes \emph{formal summaries} that are attached to difficult-to-reason-about blocks of the program, such as a particular function or loop. A formal summary is typically written in the same logic language as the actual formal specification and either partially or fully encodes the block's behavior or gives the prover hints about how to reason about the block. 
\item The user reads the \emph{final output} of the prover which ideally is a formal proof that can be mechanically checked by a \emph{proof checker} that is orders of magnitude simpler and smaller than the prover itself. Unfortunately, most systems simply output either \emph{"yes, the program adheres to the spec"} or \emph{"no, the program does not adhere to the spec"}, which introduces trust issues discussed in Section~\ref{sec:trust}.
\item  The user reads the \emph{auxiliary output} of the prover that may give the user some information about why the prover is not able to prove the spec within a reasonable time period. This auxiliary output typically requires deep understanding of the internals of the prover's implementation as well as its underlying mathematical foundations. 
\end{itemize}

\subsection{Undecidability}
\label{sec:undecidability}

The undecidability of the halting problem~\cite{Turing:1937} has a direct impact on the usability of formal verification systems. Rice's theorem~\cite{Rice:1953}, an immediate consequence of the halting problem, states that all non-trivial, semantic properties of programs are undecidable. Thus, it is impossible for a prover to be fully automatic across all programs, for any interesting semantic property. This limitation implies that a user will inevitably need to interact with the prover, in some manner beyond simply writing the formal specification, because the prover cannot, in general, be fully automatic and will need the user's help to push the proof through.

For example, provers requires that the user supplies a formal invariant for each dynamically-bound loop in a program, and these invariants are difficult for non-experts to write. For example, see the \verb|encodepacked_keccak00| (\S~\ref{sec:encodepacked-keccak00}) and \verb|encodepacked_keccak01| (\S~\ref{sec:encodepacked-keccak01}) programs, the \verb|ecrecoverloop01| (\S~\ref{sec:ecrecoverloop01}) program, and the \verb|storage02| (\S~\ref{sec:storage02}) program.

\subsection{Language Level and Compilation Stage}
A practical formal verification system will make many design decision in its implementation, but one of the most impactful is choosing the programming language and compilation stage the prover will operate at. Systems will typically take one of the following approaches:

\subsubsection{Higher Language Level / Early Compilation Stage}
With this approach, the prover is designed to work directly on the program in its original form, in the source language in which it was written, before any compilation step and without any significant source code transformations. This approach has several disadvantages:

\begin{enumerate}
\item High-level languages such as Solidity or C++ are enormous, and it would require immense engineering resources to model every language construct. For example, the Solidity language includes inline assembly which means that a prover that operates at the Solidity level would need to be engineered to reason about every inline assembly instruction as well the rest of the high level Solidity language. 
\item The prover would only work for the chosen high-level programming language and would need to be actively maintained and updated for every new version of the language, which is costly especially for a language like Solidity which is constantly being iterated on~\cite{SolidityReleases:2019}.  
\item The prover's conclusion about the correctness of the program would only apply to the high-level program and not the final binary produced by compiler which is what actually gets executed. The compiler could have a bug, for example in its optimization passes, which incorrectly changes the semantics of the program in its binary form.
\end{enumerate}

On the other hand, a very compelling advantage of this approach, which perhaps compensates for the disadvantages listed above, is that because the prover works at the language level at which the program was written, it makes it easier for the user to interact with the prover, make sense of its behavior, and understand its auxiliary output (Section~\ref{sec:user-interactions}). As explained in Section~\ref{sec:undecidability}, the user will inevitably need to interact closely with a prover because of the theoretical limits of fully automated formal verification. 

\subsubsection{Lower Language Level / Late Compilation Stage}
With this approach, the prover is designed to work after the last compilation stage, directly on the binary produced by the compiler, for example on optimized EVM bytecode or x86 instructions. This approach as several major advantages which somewhat mirror the disadvantages of operating at a higher-level:

\begin{enumerate}
\item Lower level languages such as EVM bytecode and x86 will reduce the richness of higher level language constructs to a set of basic word-level operations. For example, in Solidity, the \verb|mapping| data structure is compiled down to simple key-value pairs in an account's storage. This reduction drastically simplifies the engineering of the prover.
\item The prover's conclusion about the correctness of the program applies to the program in its executable form, which means the user does not need to trust the compiler.
\item The prover can be applied to programs originally written in any high-level language that compiles down to the same bytecode or binary language. For example, a prover that works on EVM bytecode could be used for programs originally written in both Solidity and Vyper by compiling it down to EVM bytecode first.
\end{enumerate}

However, a critical disadvantage of this approach is that the user experience of interacting with a prover that operates at the bytecode level is prohibitively tedious and time-consuming. Some examples of this phenomenon are discussed in Section~\ref{sec:walkthrough} in the context of proving properties of the \verb|SimpleMultiSig| smart contract at the EVM bytecode level.

\subsubsection{Intermediate Language Level / Mid Compilation Stage}
\label{sec:intermediate-language}
A common approach is to design the prover to operate on the intermediate language of a standard compiler infrastructure, such LLVM~\cite{Lattner:2004} or yul~\cite{Yul:2015}. An intermediate language reduces the higher level language to a more manageable set of operations and data structures, but also retains some of the important high-level constructs that make formal verification easier such as structured control flow, functions and some type information. This approach is a \emph{compromise} between strictly operating the prover at either a higher-level or lower-level language, which mitigates the impact of both their advantages and disadvantages. For example:

\begin{itemize}
\item LLVM bitcode is much more readable than x86 assembly but still much less readable that C++. Readability is important because it enables the user to interact with the prover more effectively.
\item Many higher-level languages compile down to LLVM bitcode which enables the prover to work on programs originally written in any of those languages, but now the user has to trust the portion of the compiler infrastructure that generates machine code from LLVM bitcode because the prover only operates on the LLVM bitcode.
\end{itemize}

One notable instance of this approach is to create a new, specialized intermediate programming language specifically designed for formal verification instead of machine-code generation. For example, the solc-verify~\cite{Hajdu:2019} and Verisol verification tools~\cite{Shuvendu:2019} translate Solidity programs to Boogie~\cite{Barnett:2005}, a verification intermediate language. Similar, in~\cite{Bhargavan:2016}, Solidity programs are translated to F*.

\subsection{Formalization of Correctness Properties}
The first step to formally verifying a program is informally, yet rigorously, describing what the intended correct behavior of the program is, and then translating this informal description into a formal specification that a mechanized prover can understand. This translation step is critical because the proof generated by the prover assumes that the formal specification is correct in the sense that it faithfully captures the intended correctness properties of the program.

\subsubsection{Formal specifications need be simple for humans to read and understand}
The premise of formal verification is that formal specifications are much simpler to understand and be convinced of being correct than the program. It is unclear how useful formal verification is when the formal specification is longer or more difficult to read than the program itself.

\subsubsection{A partial versus a full formal specification}
\label{sec:partial-versus-full}
Users typically do not write formal specifications for all correctness properties of a program because writing and proving formal specifications is time-consuming and difficult. Instead, in practice, users consider the cost and benefit of formally specifying a property. See Section~\ref{sec:bytes00} for an example of making this practical trade-off.

\subsection{Provers are not trustworthy}
\label{sec:trust}

A prover often consists of hundreds of thousands of lines of highly complex code and thus bugs in the prover are inevitable which could cause the prover to incorrectly claim that a program follows specification when it does not. A technique called proof carrying code~\cite{Necula:1997} has been extensively researched in academia that aims to address this challenge, but has not been used in practice. 

\subsection{Investigating Why the Prover Fails}
\label{sec:prover-debugging}
The prover may fail to prove a specification for several reasons: 

\begin{enumerate}
\item The specification has a bug.
\item The program has a bug.
\item The prover has a bug.
\item The language semantics has a bug.
\item The underlying constraint solver (e.g Z3~\cite{Demoura:2008}) has a bug.
\item An incorrect lemma was supplied.
\item The prover is not powerful enough to reason automatically about the program and specification.
\end{enumerate}

Trying to figure why the prover fails is tedious, difficult and time-consuming because typically a user would need to understand the internals of the prover implementation as well as its method of deductive reasoning. Conceptually, a prover works by carefully exploring the entire, exponentially-sized state space of the program under \emph{all} possible inputs. Some tools such as KLab~\cite{KLab:2018} have been developed to help users navigate this state space, similar to how a debugger can be used to step through the state changes of a program under a single input.


\subsection{EVM Bytecode}
Some of the design of EVM bytecode makes formal verification tedious, difficult, and time-consuming. For example, the EVM does not have functions, which means that a user will need to specify the actual program counter of the EVM bytecode that the specification applies to. If the user changes the Solidity program and recompiles it to EVM bytecode, the user will need to manually update the program counter in the specification. This manual process is error-prone because if the user gets it wrong, the specification will be incorrect and thus a buggy program may pass through undetected. 



\section{K-YAML}
\label{sec:kyaml}
The K specification language is based on matching logic~\cite{Rosu:2017} which makes it powerful, expressive, and flexible, but its \emph{syntax} can be difficult for new users to read. K's syntax was originally designed to make it easier for formal methods experts to specify complete executable semantics for entire programming languages, and thus, one of its design priorities was optimizing the succinctness of writing individual rules. However, for programmers without experience in formal methods who only want to use K to write correctness specifications for their programs, the syntax can initially seem prohibitively esoteric. 

This paper proposes a simple YAML-based format for structuring K specifications called K-YAML that enables users to write easier-to-read specifications for formally verifying programs. YAML was chosen because it is a widely-used, human-readable, system configuration file format that programmers are comfortable with. The key design decision of K-YAML is to be purely syntactic sugar for K and to not abstract away any part of it in order to retain its power and expressivity.

A K-YAML specification is a list of \verb|spec| blocks where each block has the following structure:
\begin{lstlisting}[language=KYaml]
spec:
  - rule:
      name: <Name-of-Spec>
      inherits: <Name-of-Parent-Spec>
      if:
        match:
          <If-Terms>
        where:
          <If-Constraints>
      then:
        match:
          <Then-Terms>
        where:
          <Then-Constraints>
\end{lstlisting}
\begin{itemize}
    \item The \verb|name| key designates a name for this \verb|spec| block that can be referenced elsewhere in the specification.
    \item The  \verb|inherits| key references another \verb|spec| block to indicate that this one inherits from it. (Section~\ref{sec:requires00})
    \item The \verb|if| key is the \emph{precondition}, which defines the set of initial program states of interest using two components: \verb|match| and \verb|where|. Conceptually, the \verb|match| component is similar to Rust's \verb|match| operator which accepts patterns over terms and variables to describe the structure of data and then matches and binds a value against the structure. Here, \verb|match| is a dictionary whose keys are K configuration cells and whose values are K terms over symbolic variables. The \verb|where| component is a list of conjuncts over K predicates and symbolic variables that have possibly been bound by the \verb|match| component. A program state is in the precondition if it matches the \verb|match| dictionary and satisfies the \verb|where| constraints. 
    \item The \verb|then| key is the \emph{postcondition} which specifies what the final program states should be in terms of the initial program states defined by the precondition. It uses the same \verb|match| and \verb|where| mechanism to specify this set of final program states.
\end{itemize} 
Section~\ref{sec:walkthrough} gives many examples of small programs and the corresponding correctness specifications in K-YAML.

\section{Verification Walkthrough}
\label{sec:walkthrough}
This section details some of my experience using the K prover to verify a few properties of the \verb|SimpleMultiSig| smart contract. I was not involved in the K prover's design and implementation, and thus, I needed to find a way to use it effectively without understanding its internals. My approach was to start by formally verifying simple programs and then incrementally adding new functionality until I reconstructed the \verb|SimpleMultiSig|. This approach enabled me to use the prover as a black box as much as possible, thus minimizing the need to understand its internals. However, I did need to have a thorough understanding of the K-EVM semantics~\cite{Hildenbrandt:2018} to make progress.

\subsection{Starting Point}
\label{sec:simple00}

As a starting point, I used the simple program below which has a single function named \verb|execute| that returns the value 5:

\begin{lstlisting}[language=Solidity]
pragma solidity 0.5.0;
contract simple00 {
    function execute() public returns (uint) {
        return 5;
    }
}
\end{lstlisting}

Then I wrote the specification below to try to prove that each transaction to the \verb|execute| function always returns the value 5 successfully. More technically stated, the specification ensures that when the calldata of a transaction matches the ABI-encoding of \verb|execute|’s function-selector, the value returned is the ABI-encoding of the constant 5 as a 32-byte \verb|uint256|, and the status code of the transaction is \verb|EVMC_SUCCESS|:
 
\begin{lstlisting}[language=KYaml]
spec:
  - rule:
      if:
        match:
          callData: \#abiCallData("execute", .TypedArgs)
      then:
        match:
          statusCode: EVMC_SUCCESS
          output: \#encodeArgs(#uint256(5))
\end{lstlisting}

Then I compiled the program to EVM bytecode using the Solidity compiler and invoked the K prover on the bytecode and the specification. After a few seconds, the prover returned \verb|True|, which indicates that it was able to formally prove that the program passes the specification for all possible transactions, in the context of all possible blockchain states.


\subsection{Using function parameters}
\label{sec:simple02}

Next, I took the previous program and modified it by 1) adding a parameter \verb|a0| to the \verb|execute| function and 2) returning \verb|a0| instead of the constant 5. 

\begin{lstlisting}[language=Solidity]
pragma solidity 0.5.0;
contract simple02 {
    function execute(uint a0) public returns (uint) {
        return a0;
    }
}
\end{lstlisting}

The specification had to be modified as follows:
\begin{enumerate}
    \item The \verb|callData| cell was changed by introducing a fresh symbolic variable \verb|A0| that is bound to the first, 32-byte argument of the calldata. 
    \item The output cell was changed to return \verb|A0|, thus specifying that \verb|execute| returns the value of its first, 32-byte argument. 
    \item Finally, a \verb|where| constraint is added to constrain the value of the symbolic variable \verb|A0| to be within the \verb|uint256| domain, which includes all integers between $0 \leq A0 < 2^{256}$.
\end{enumerate}

\begin{lstlisting}[language=KYaml]
spec:
  - rule:
      if:
        match:
          callData: \#abiCallData("execute",\#uint256(A0))
        where:
          - \#rangeUInt(256, A0)
      then:
        match:
          statusCode: EVMC_SUCCESS
          output: \#encodeArgs(\#uint256(A0))
\end{lstlisting}

\subsection{Static Arrays}
\label{sec:staticarray00}

Next, I changed the program to accept a static array parameter and return its first element.

\begin{lstlisting}[language=Solidity]
pragma solidity 0.5.0;
contract staticarray00 {
    function execute(uint[3] memory a)
            public returns (uint) {
        return a[0];
    }
}
\end{lstlisting}

Because the KEVM semantics did not have an existing a high-level construct for expressing static array calldata parameters, I needed to add it myself. Fortunately, the ABI encoding of statically-sized data structures is simple: they are straightforwardly flattened into a tuple of 32-byte elements. I defined a new K rule named \verb|#abiCallData2| that behaves exactly like the existing \verb|#abiCallData| function except that it enables the user to specify the full function signature instead of only the function name: 

\begin{lstlisting}[language=K]
syntax WordStack ::= 
    #abiCallData2 ( String , TypedArgs ) [function]

rule #abiCallData2( FSIG , ARGS )
 	=> #parseByteStack(substrString(Keccak256(FSIG), 0, 8))
 	++ #encodeArgs(ARGS)
\end{lstlisting}

This new \verb|#abiCallData2| function enabled me to represent the static array parameter \verb|a| as the tuple \verb|(A0, A1, A2)| where \verb|A0|, \verb|A1|, \verb|A2| are 32-byte, fresh symbolic variables:

\begin{lstlisting}[language=KYaml]
spec:
  - rule:
      if:
        match:
          callData: \#abiCallData2("execute(uint256[3])",
            \#uint256(A0), \#uint256(A1), \#uint256(A2))
        where:
          - \#rangeUInt(256, A0)
          - \#rangeUInt(256, A1)
          - \#rangeUInt(256, A2)
      then:
        match:
          statusCode: EVMC_SUCCESS
          output: \#encodeArgs(\#uint256(A0))
\end{lstlisting}

This is a good example of the flexibility and extensibility of K compared to other verification systems: extensions can be incorporated without needing to make changes to the underlying prover. The new K rule above gave the abstraction needed to succinctly write the spec.

\subsection{Dynamic byte arrays}
\label{sec:bytes00}

This next program accepts a dynamically-sized, \verb|bytes|-typed parameter named \verb|data| and return its length:

\begin{lstlisting}[language=Solidity]
pragma solidity 0.5.0;
contract bytes00 {
    function execute(bytes memory data)
            public returns (uint) {

        return data.length;
    }
}
\end{lstlisting}

The specification follows the same pattern as the previous specifications but  uses the KEVM \verb|#buf| construct to bind dynamically-sized calldata:

\begin{lstlisting}[language=KYaml]
spec:
  - rule:
      if:
        match:
          callData: \#abiCallData2("execute(bytes)",
                      \#bytes(\#buf(DATA_LEN,DATA)))
        where:
          - \#range(0 <= DATA_LEN < 2 ^Int 16)
      then:
        match:
          statusCode: EVMC_SUCCESS
          output: \#encodeArgs(\#uint256(DATA_LEN))
\end{lstlisting}

The \verb|#buf(DATA_LEN,DATA)| expression is a symbolic byte buffer that introduces two fresh symbolic variables to bind the \verb|data| parameter. \verb|DATA_LEN| represents \verb|data|'s length and \verb|DATA| represents its contents. 

Note the \verb|where| constraint in the specification which bounds the length of \verb|DATA_LEN| to be between $0$ and $2^{16}$. This constraint is needed because the EVM imposes a $2^{32}$ upper limit on the length of a \verb|bytes| array since it must fit into memory, which is 32-byte addressable. However, this $2^{32}$ upper bound is actually not tight enough to make the specification pass because the EVM will need to store other data in memory besides the \verb|bytes| array. I could not figure out a quick way to calculate the tightest possible upper bound on \verb|DATA_LEN|, so I did what was done for other K verification projects such as the GnosisSafe~\cite{RV:2019}, which was to arbitrarily constrain \verb|DATA_LEN| to $2^{16}$.

Thus, technically-speaking, the correctness guarantees given by the formal proof would not apply to transactions whose data parameter is longer than $2^{16}$ bytes. Practically speaking, it is unclear whether it would be worth the effort to figure out the tightest upper bound. This situation is an example of the common trade off between the precision and cost of partial versus full formal verification, as discussed in Section~\ref{sec:partial-versus-full}.

This is a good example of three usability challenges: automation (Section~\ref{sec:undecidability}), low-level specifications (Section~\ref{sec:intermediate-language}), and prover debugging (Section~\ref{sec:prover-debugging}). The user needs to understand the EVM at a low level to write a passing specification, and the actual tight bound to make this specification pass is hard for a user to calculate. Furthermore, the user’s first attempt at writing this specification would likely be to erroneously constrain \verb|DATA_LEN| to $2^{32}$ and when the prover fails to prove this, it is very non-obvious why and the output of the K prover does not help because it only gives the set of all intermediate proof states. It is possible that, if the prover implementation was able to generate a counter-example, that is, a concrete transaction that demonstrates the program violating the specification, the user may have a better chance of figuring out the reason for the specification failing.

\subsection{Transaction Reverts}
\label{sec:requires00}

This next program uses Solidity's \verb|require| statement which reverts the transaction when the \verb|a0 > 0| condition does not hold:
\begin{lstlisting}[language=Solidity]
pragma solidity 0.5.0;
contract requires00 {
    function execute(uint256 a0)
            public returns (uint256) {

        require(a0 > 0);
        return 5;
    }
}

\end{lstlisting}

The specification now needs to cover more than one program path: a success path when $a0 > 0$ and revert path when $\neg a0 > 0$. Thus it has two \verb|spec| blocks, one for each path:

\begin{lstlisting}[language=KYaml]
- spec:
    name: a0gt0
    if:
      match:
        callData: #abiCallData2("execute(uint256)", #uint256(A0))
      where:
        - A0 >Int 0
        - #rangeUInt(256, A0)
    then:
      match:
        output: #encodeArgs(#uint256(5))
        statusCode: EVMC_SUCCESS

- spec:
    name: a0le0
    if:
      match:
        callData: #abiCallData2("execute(uint256)", #uint256(A0))
      where:
        - A0 <=Int 0
        - #rangeUInt(256, A0)
    then:
      match:
        statusCode: EVMC_REVERT
\end{lstlisting}

The first \verb|spec| named \verb|a0gt0| specifies the $a0 > 0$ case: the status code of the execute function is \verb|EVMC_SUCCESS| and the output is the constant 5. The second \verb|spec| named \verb|a0le0| specifies the $\neg a0 > 0$ case: the status code of the execute function is \verb|EVMC_REVERT| and the output is unspecified.

The specification can be slightly rewritten so that common parts of the two \verb|spec| blocks can be shared, instead of duplicated, in a \emph{parent} \verb|spec| by using the \verb|inherits| property:

\begin{lstlisting}[language=KYaml]
spec:
  - rule:
      name: base
      if:
        match:
          callData: \#abiCallData2("execute(uint256)", \#uint256(A0))
        where:
          - \#rangeUInt(256, A0)

  - rule:
      name: a0gt0
      inherits: base
      if:
        where:
          - A0 >Int 0
      then:
        match:
          output: \#encodeArgs(\#uint256(5))
          statusCode: EVMC_SUCCESS

  - rule:
      name: a0le0
      inherits: base
      if:
        where:
          - A0 <=Int 0
      then:
        match:
          statusCode: EVMC_REVERT
\end{lstlisting}

The rewritten specification above introduces a third block named \verb|base| that lifts the common parts, and the \verb|a0gt0| and \verb|a0le0| blocks now \verb|inherit| the \verb|base| block. While it may not seem that this refactoring makes the specification more concise in this particular example, when specifying larger programs, this feature will be necessary to keep the specification readable as can be seen in final \verb|SimpleMultiSigT3| specification in Section~\ref{sec:specification}.

\subsection{Hashing packed encoding}
\label{sec:encodepacked-keccak00}
\label{sec:encodepacked-keccak01}

This next program calculates and returns the keccak-256 cryptographic hash  of a \verb|bytes| array parameter prefixed with the single byte of 0x1:

\begin{lstlisting}[language=Solidity]
pragma solidity 0.4.24;
contract encodepacked_keccak01 {
    function execute(bytes32 a0)
            pure external returns(bytes32) {
        return keccak256(
                    abi.encodePacked(byte(0x01),a0));
    }
}
\end{lstlisting}

The specification uses the KEVM semantics function named \verb|keccak256| that abstracts the keccak-256 cryptographic hash function as an uninterpreted function that assumes all the hashed values appearing in each execution trace are collision-free:

\begin{lstlisting}[language=KYaml]
- spec:
    if:
      match:
        callData:  #abiCallData2("execute(bytes32)",
                    #bytes32(A0))
      where:
        - #rangeUInt(256, A0)
    then:
      match:
        output: #encodeArgs(#bytes32(keccak(1 :
                  #encodeArgs(#uint256(A0)))))
        statusCode: EVMC_SUCCESS
\end{lstlisting}

This abstraction of keccak-256 is the typical approach taken by practical formal verification systems, where complex library functions are abstracted by writing a formal specification for their behavior called a \emph{formal summary}, and then applying the formal summary at call sites invoking the library function instead of directly verifying the library function's code. This method of abstraction enables the user to decompose the verification task into two steps: first formally verifying the program that uses the library function, and then separately verifying the implementation of the library function. See Section~\ref{sec:ecrecover00} for another example of how this formal summary technique is used for programs that verify signatures.

Now, when I tried to run the prover, I found that it failed to prove the specification even though the specification is correct. It turned out that the reason is that the actual EVM bytecode that this Solidity program compiles down to has a loop that is essentially an inlined "memcpy" for copying calldata to local memory. As discussed in Section~\ref{sec:undecidability}, in general, provers will not be able to reason about loops with dynamic bounds automatically unless the user supplies superfluous information such as a loop invariant. 

The figure below illustrates this memcpy loop in the control flow graph of the EVM bytecode of the \verb|encodepacked_keccak00| program. 

\includegraphics[width=.9\linewidth]{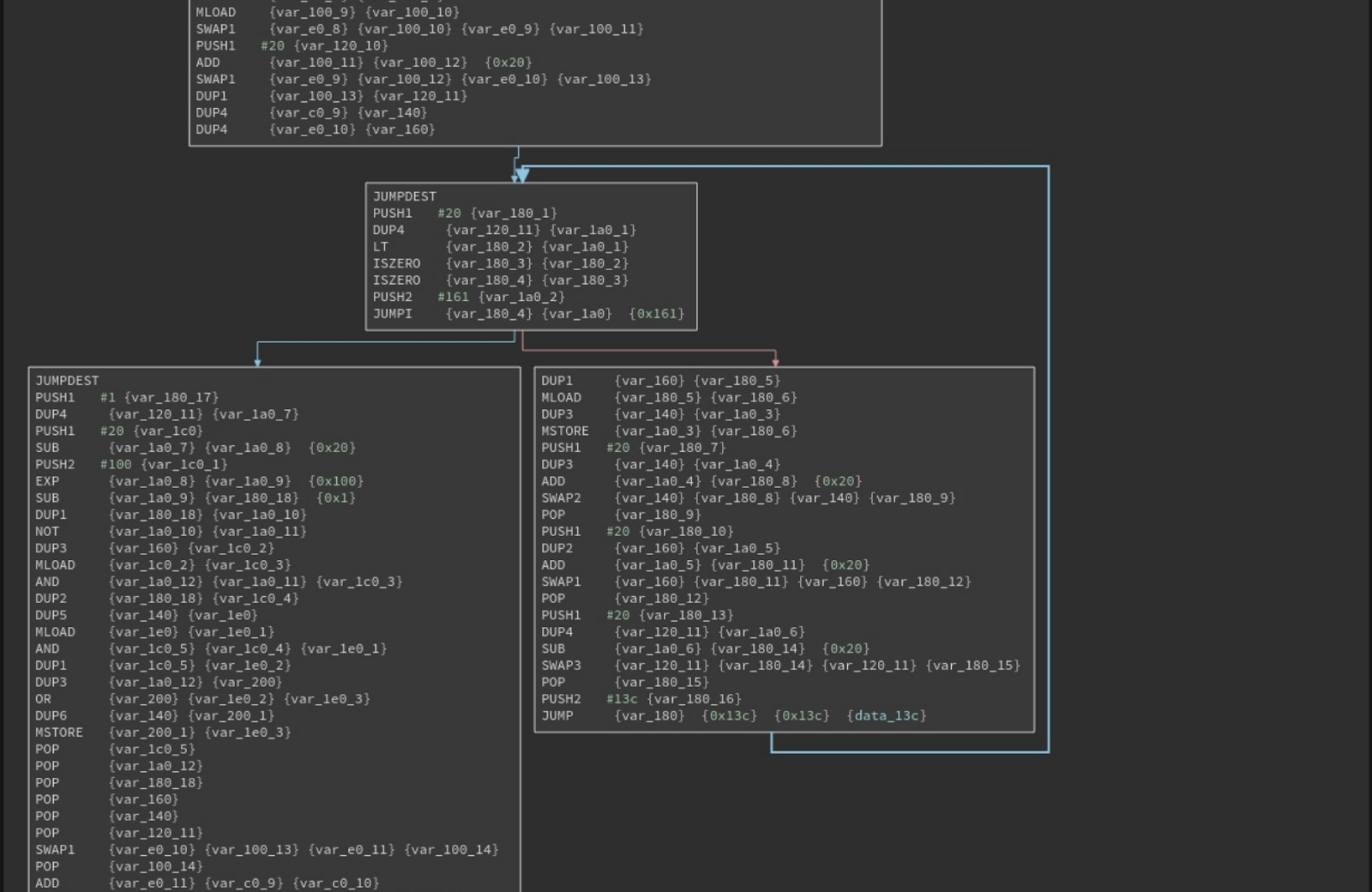}

I could have spent time writing a loop invariant but fortunately, in this case, there’s a simpler approach which was to update to the 0.5.0 version of the Solidity compiler instead of 0.4.24 because 0.5.0 has optimizations that are able to optimize away the loop entirely. The following program, which has been changed to use \verb|pragma solidity 0.5.0|, has an acyclic control flow graph and the prover is able verify it.

\begin{lstlisting}[language=Solidity]
pragma solidity 0.5.0;
contract encodepacked_keccak01 {
    function execute(bytes32 a0)
            pure external returns(bytes32) {
        return keccak256(
            abi.encodePacked(byte(0x01),a0));
    }
}
\end{lstlisting}

This program is a good example of several usability challenges that formal verification systems typically exhibit: automation, low-level specifications, and prover debugging. It would be difficult for a user to figure out that the compiled EVM bytecode has a hidden loop that does not appear in the original Solidity source program. If the prover was fully automated, this hidden loop would not be an issue because the prover would just be able to prove the specification without user intervention. But, because it cannot be fully automated due to undecidability, the user needs to supply the prover with a loop invariant in terms of the compiled EVM bytecode, which means the user will have to decompile the bytecode, identify the exact instruction counters of the loop entry/exit and make sense of the difficult-to-follow stack-based push/pop/dup instructions of the EVM.

\subsection{Statically-sized loops}
\label{sec:staticloop00}

This next program has a loop that iterates three times after an initial \verb|require| check:

\begin{lstlisting}[language=Solidity]
pragma solidity 0.5.0;
contract staticloop00 {
    function execute(uint a0)
            pure external returns(uint256) {
        uint sum = a0;
        require (a0 < 10);
        for (uint i = 0; i < 3; i++) {
            sum += i;
        }
        return sum;
    }
}
\end{lstlisting}

Note that statically-bound loops, such as this one, are much easier for a prover to reason about automatically than a dynamically-bound loop. In principle, a prover can automatically and fully unroll a statically-sized loop into straight-line code. In contrast, dynamically-bound loops require that the user supply a loop-invariant.

The specification below checks that, if the function parameter is an integer within [0, 10), the transaction will return the value of $a0+3$ with an \verb|EVMC_SUCCESS| status code. Otherwise, it will revert with a \verb|EVMC_REVERT| status code:

\begin{lstlisting}[language=KYaml]
spec:
  - rule:
      if:
        match:
          callData: \#abiCallData2("execute(uint256)", \#uint256(A0))
        where:
          - \#range(0 <= A0 < 10)
      then:
        match:
          statusCode: EVMC_SUCCESS
          output: \#encodeArgs(\#uint256(A0 +Int 3))
\end{lstlisting}

\subsection{Recovering an address from a signature}
\label{sec:ecrecover00}

This next program accepts a 32-byte message and the (V,R,S) components of an ECDSA signature. It uses the Solidity \verb|ecrecover| function to verify the signature on the message and recover the signer address, and otherwise reverts:

\begin{lstlisting}[language=Solidity]
pragma solidity 0.5.0;
contract ecrecover00 {
    function execute(bytes32 hash, uint8 sigV,
                     bytes32 sigR, bytes32 sigS)
                     pure external returns(address) {
        address a = ecrecover(hash, sigV, sigR, sigS);
        require(a > address(0));
        return a;
    }
}
\end{lstlisting}

The specification below has three blocks: 1) the \verb|sigvalid| block specifies that the behavior of the program on a transaction with a valid signature is to return the signer address, 2) the \verb|siginvalid| block specifies that it should revert if the signature is not valid, and 3) the \verb|base| block is a parent of the other two blocks that keeps the common components. 

\begin{lstlisting}[language=KYaml]
spec:
  - rule:
      name: base
      if:
        match:
          callData: \#abiCallData2("execute(bytes32,uint8,bytes32,bytes32)",
              \#bytes32(HASH), \#uint8(SIGV),
              \#bytes32(SIGR), \#bytes32(SIGS))
        where:
          - \#rangeUInt(256, HASH)
          - \#rangeUInt(8, SIGV)
          - \#rangeBytes(32, SIGR)
          - \#rangeBytes(32, SIGS)
          - ECREC_DATA ==K
              \#encodeArgs(\#bytes32(HASH), \#uint8(SIGV),
                 \#bytes32(SIGR), \#bytes32(SIGS))

  - rule:
      name: sigvalid
      inherits: base
      if:
        where:
          - RECOVERED ==Int \#symEcrec(ECREC_DATA)
          - not: \#ecrecEmpty(ECREC_DATA)
      then:
        match:
          statusCode: EVMC_SUCCESS
          output: \#encodeArgs(\#address(RECOVERED))

  - rule:
      name: siginvalid
      inherits: base
      if:
        where:
          - \#ecrecEmpty(ECREC_DATA)
      then:
        match:
          statusCode: EVMC_REVERT
          output: _
\end{lstlisting}

The KEVM semantics defines two functions named \verb|#symEcrec| and \verb|#ecrecEmpty| that abstract the \verb|ecrecover| precompile as an uninterpreted function, similar to how the keccak-256 cryptographic hash function is abstracted (Section~\ref{sec:encodepacked-keccak00}).

\subsection{Recovering a static array of signatures}
\label{sec:ecrecoverloop00}
\label{sec:ecrecoverloop01}
This next program uses a static loop (Section~\ref{sec:staticloop00}) to iterate over a static array (Section~\ref{sec:staticarray00}) of signatures (Section~\ref{sec:ecrecover00}) and reverts (Section~\ref{sec:requires00}) if any of them do not verify. Otherwise, the function returns successfully. This program uses four different constructs that were previously verified independently of each other in earlier sections, but now they are combined into a single program. This is a good example of how a user can incrementally build up the specification of their program without needing to understand the internals of how the prover works.
  
\begin{lstlisting}[language=Solidity]
pragma solidity 0.5.0;
contract ecrecoverloop00 {
    function execute(bytes32 hash,
                     uint8[2] memory sigV,
                     bytes32[2] memory sigR,
                     bytes32[2] memory sigS)
            pure public {

        for (uint i = 0; i < 2; i++) {
            address a = ecrecover(hash, sigV[i],
                                    sigR[i], sigS[i]);
            require(a > address(0));
        }
    }
}
\end{lstlisting}

The specification has four blocks. The \verb|base| block serves as the prefix of the three other blocks and specifies the behavior related to function entry such calldata. The \verb|sigs-valid| block specifies the success case where all signatures can be verified. The \verb|sigs0-invalid| and \verb|sigs1-invalid| blocks specify the cases where the first and second signatures, respectively, cannot be verified:

\begin{lstlisting}[language=KYaml]
spec:
  - rule:
      name: base
      if:
        match:
          callData: \#abiCallData2("execute(bytes32,uint8[2],bytes32[2],bytes32[2])",
                      \#bytes32(HASH),
                      \#uint8(SIGV0), \#uint8(SIGV1),
                      \#bytes32(SIGR0), \#bytes32(SIGR1),
                      \#bytes32(SIGS0), \#bytes32(SIGS1))
        where:
          - \#rangeUInt(256, HASH)
          - \#rangeUInt(8, SIGV0)
          - \#rangeUInt(8, SIGV1)
          - \#rangeBytes(32, SIGR0)
          - \#rangeBytes(32, SIGR1)
          - \#rangeBytes(32, SIGS0)
          - \#rangeBytes(32, SIGS1)
          - ECREC_DATA0 ==K
              \#encodeArgs(\#bytes32(HASH), \#uint8(SIGV0),
              \#bytes32(SIGR0), \#bytes32(SIGS0))
          - ECREC_DATA1 ==K
              \#encodeArgs(\#bytes32(HASH), \#uint8(SIGV1),
              \#bytes32(SIGR1), \#bytes32(SIGS1))

  - rule:
      name: sigs-valid
      inherits: base
      if:
        where:
          - RECOVERED0 ==Int \#symEcrec(ECREC_DATA0)
          - RECOVERED1 ==Int \#symEcrec(ECREC_DATA1)
          - not: \#ecrecEmpty(ECREC_DATA0)
          - not: \#ecrecEmpty(ECREC_DATA1)
      then:
        match:
          statusCode: EVMC_SUCCESS

  - rule:
      name: sig0-invalid
      inherits: base
      if:
        where:
          - \#ecrecEmpty(ECREC_DATA0)
      then:
        match:
          statusCode: EVMC_REVERT

  - rule:
      name: sig1-invalid
      inherits: base
      if:
        where:
          - \#ecrecEmpty(ECREC_DATA1)
      then:
        match:
          statusCode: EVMC_REVERT
\end{lstlisting}

The prover was able to prove the specification above. But, then, I made a slight modification to the \verb|ecrecoverloop00| program by only adding a new \verb|bytes|-typed parameter named \verb|data|, which is not used at all in the body of the Solidity function:

\begin{lstlisting}[language=Solidity]
pragma solidity 0.5.0;
contract ecrecoverloop01 {
    function execute(bytes32 hash, bytes memory data,
                     uint8[2] memory sigV,
                     bytes32[2] memory sigR,
                     bytes32[2] memory sigS)
            pure public {

        for (uint i = 0; i < 2; i++) {
            address a = ecrecover(hash, sigV[i],
                                    sigR[i], sigS[i]);
            require(a > address(0));
        }
    }
}
\end{lstlisting}

To clarify, the reason I added this dead parameter is that my goal is to eventually reconstruct the \verb|SimpleMultiSig| program which accepts such a \verb|bytes| parameter in the same function that verifies signatures using a loop.

Unintuitively, simply adding this \emph{dead} \verb|data| parameter causes the prover to timeout on the \verb|sig1-invalid| specification, when using a time limit of 30 minutes. In contrast, without this parameter, the prover returns successfully within three minutes. It turned out that, in order to fix this issue, I needed to add the additional, \emph{superfluous} \verb|where| constraint \verb|not: #ecrecEmpty(ECREC_DATA0)| on line 6 to the \verb|sig1-invalid| specification:

\begin{lstlisting}[language=KYaml]
- spec:
    name: sig1-invalid
    inherits: base
    if:
      where:
        - not: #ecrecEmpty( ECREC_DATA0 )
        - #ecrecEmpty(ECREC_DATA1)
    then:
      match:
        statusCode: EVMC_REVERT
\end{lstlisting}

This additional constraint is superfluous in the logical sense because it does not change the specification’s logical meaning since $A \lor B$ is equivalent to $A \lor (\neg A \land B)$. Conceptually, adding this additional conjunct helps the prover succeed because it directly captures the behavior of the for-loop. There may exist multiple invalid signatures, but the for-loop terminates when it reaches the very first invalid signature, so the specification needed to faithfully capture this behavior.

\subsection{Reading from storage}
\label{sec:storage00}

This next program reads the storage variable \verb|n| and returns its value:

\begin{lstlisting}[language=Solidity]
pragma solidity 0.5.0;
contract storage00 {
    uint private n;

    function execute() view public returns(uint) {
        return n;
    }
}
\end{lstlisting}

The corresponding specification below shows how the KEVM semantics abstracts storage as an integer map represented by fresh symbolic variable \verb|S| and uses the KEVM \verb|select| function to calculate the  offset per Solidity's interpretation of the EVM storage:

\begin{lstlisting}[language=KYaml]
spec:
  - rule:
      if:
        match:
          storage: S
          callData: \#abiCallData2("execute()", .TypedArgs)
        where:
          - N ==Int select(S, \#hashedLocation("Solidity",
                                    0, .IntList))
          - \#rangeUInt(256, N)
      then:
        match:
          output: \#encodeArgs(\#uint256(N))
          statusCode: EVMC_SUCCESS
\end{lstlisting}

\subsection{Writing to storage}
\label{sec:storage01}

This next program writes the constant 5 to the storage variable \verb|n|, reads \verb|n| and then returns its value:

\begin{lstlisting}[language=Solidity]
pragma solidity 0.5.0;
contract storage01 {
    uint private n;

    function execute() public returns(uint) {
        n = 5;
        return n;
    }
}
\end{lstlisting}

The EVM will give a \emph{refund} of ether to the caller if a transaction writes the value of 0 to a storage location, which enables the EVM to stop explicitly tracking that storage location on-chain. The rationale of this refund mechanism is to incentivize users to free up storage they are not using anymore. For this specific program, however, there is never a refund because storage is always written with the value 5, not 0. Thus, the specification below has the \verb|refund| cell set to 0 indicating a caller will never receive a refund under any transaction.

\begin{lstlisting}[language=KYaml]
spec:
  - rule:
      if:
        match:
          storage: S0
          refund: 0
          callData: \#abiCallData2("execute()", .TypedArgs)
        where:
          - N0 ==Int select(S0, \#hashedLocation("Solidity",
                                    0, .IntList))
          - \#rangeUInt(256, N0)
      then:
        match:
          storage: S1
          output: \#encodeArgs(\#uint256(N1))
          statusCode: EVMC_SUCCESS
        where:
          - N1 ==Int select(S1, \#hashedLocation("Solidity",
                                    0, .IntList))
          - N1 ==Int 5
\end{lstlisting}

The prover should have been able prove the specification above, however, it was not able to do. Interestingly, the prover constructed a symbolic expression that had a value of zero, but it was unable to \emph{simplify} this expression to zero on its own. To fix this, I needed to add a new lemma that helped the prover understand that the expression does in fact simplify to zero:

\begin{lstlisting}[language=K]
rule Rsstore(BYZANTIUM, NEW, CURR, ORIG) => 0 requires NEW =/=Int 0
\end{lstlisting}

\subsection{Writing overflow to storage}
\label{sec:storage02}

This next program increments the storage variable \verb|n| which will overflow when \verb|n| is $2^{256} -1$:
\begin{lstlisting}[language=Solidity]
pragma solidity 0.5.0;
contract storage02 {
    uint private n;

    function execute() public returns(uint) {
        n = n + 1;
        return n;
    }
}
\end{lstlisting}
The specification needs to handle the non-overflow and overflow cases separately because the \verb|+Int| operator used by the KEVM semantics is integer addition, rather than  modular addition:

\begin{lstlisting}[language=KYaml]
spec:
  - rule:
      name: base
      if:
        match:
          callData: \#abiCallData2("execute()", .TypedArgs)
          storage: S0
          refund: _
        where:
          - N0 ==Int select(S0, \#hashedLocation("Solidity", 0, .IntList))
      then:
        match:
          storage: S1
          refund: _
          statusCode: EVMC_SUCCESS
          output: \#encodeArgs(\#uint256(N1))

  - rule:
      name: no-overflow
      inherits: base
      if:
        where:
          - \#range(0 <= N0 < 2 ^Int 256 -Int 1)
          - N1 ==Int N0 +Int 1

  - rule:
      name: overflow
      inherits: base
      if:
        where:
          - N0 ==Int 2 ^Int 256 -Int 1
          - N1 ==Int 0
\end{lstlisting}

The prover was able to prove the \verb|no-overflow| specification automatically without any user intervention, but not the \verb|overflow| specification. The reason was that \verb|output| cell of the final proof state had the term
\[\verb|#buf(32, chop(select(S0, 0) + 1))|\] 
which represents a 32-byte buffer whose value is one plus the zeroth storage slot under modular addition. The final proof state also had the constraint 
\[\verb|select(S0,0)| = 2^{256}\]
which constrains the zeroth storage slot to $2^{256}$. Ideally, the prover should have been able to simplify \verb|chop(select(S0, 0) + 1)| to the value 0 on its own, but I needed to add the following lemma which simply states that applying the \verb|chop| function to $2^{256}$ should simplify to 0:
\begin{lstlisting}[language=K]
rule chop(I) => 0 requires I ==Int pow256
\end{lstlisting}

\subsection{Calling an address}
\label{sec:call00}

This next program invokes the \verb|destination| parameter with a call payload passed by the caller:

\begin{lstlisting}[language=Solidity]
pragma solidity 0.5.0;
contract call00 {
    function execute(bool condition, uint gasLimit,
                     uint value, bytes memory data,
                     address destination) public {
        require(condition);
        bool success = false;
        assembly {
            success := call(gasLimit, destination,
                            value, add(data, 0x20),
                            mload(data), 0, 0)
        }
        require(success);
    }
}
\end{lstlisting}

The specification checks the following correctness properties:

\begin{itemize}
\item Call to \verb|destination| is invoked at least once if \verb|condition| is true

\item Call to \verb|destination| is not invoked if \verb|condition| is false

\item Call to \verb|destination| is invoked at most once if \verb|condition| is true

\item Call to \verb|destination| is invoked with the correct value, gas limit, return start/length

\item Call to \verb|destination| is invoked with correct data start/length

\item execute reverts if call to \verb|destination| returns an error code execute succeeds if call to \verb|destination| returns a success code

\item No storage reads or writes after the call to \verb|destination|.
\end{itemize}

\begin{lstlisting}[language=KYaml]
spec:
  - rule:
      name: base
      if:
        match:
            callData:  \#abiCallData("execute",
                \#bool(CONDITION),
                \#uint256(GAS_LIMIT), \#uint256(VALUE),
                \#bytes(\#buf(DATA_LEN,DATA)),
                \#address(DESTINATION))
            callLog: .List
        where:
            - \#rangeAddress(DESTINATION)
            - \#rangeUInt(8, CONDITION)
            - \#rangeUInt(256, GAS_LIMIT)
            - \#rangeUInt(256, VALUE)
            - \#rangeUInt(256, DATA_LEN)
            - DATA_LEN <Int 2 ^Int 16
            - CALL_PC ==Int 364
            - CONDITION ==Int 1
      then:
        match:
          callLog: ListItem ( 0 CALL_PC GAS_LIMIT
                      DESTINATION VALUE ARGSTART_C
                      ARGWIDTH_C 0 0 LM_C .List .List )
                   .List
        where:
          - selectRange(LM_C, ARGSTART_C, ARGWIDTH_C) ==K
                    \#buf(DATA_LEN,DATA)

  - rule:
      name: call-succeeded
      inherits: base
      if:
        where:
          - \#callSuccess(0, DESTINATION)
      then:
        match:
          statusCode: EVMC_SUCCESS

  - rule:
      name: call-failed
      inherits: base
      if:
        where:
          - \#callFailure(0, DESTINATION)
      then:
        match:
          statusCode: EVMC_REVERT
\end{lstlisting}

To check these properties, I needed to extend the KEVM semantics by adding a new cell in the KEVM configuration named \verb|callLog| to keep track of a list of all the call invocations during transaction execution. Each element in the list is a tuple representing callsite information including the call index, program counter, gas limit, memory offset and size of the parameters and return value. I also added two new cells named \verb|readLog| and \verb|writeLog| that keep track of all the storage reads and writes to check that none occur after call instructions. The details are omitted here due to space limitations.

\section{SimpleMultiSig}
\label{sec:simplemultisig}
Multisig wallets are compelling targets for formal verification because they control high-value assets.  The \verb|SimpleMultiSig| smart contract~\cite{Lundqvist:2015, Lundqvist:2017} is a minimal multisig wallet written in Solidity for the EVM. Its code is included in the appendix under Listing~\ref{lst:simplemultisig}. 


\subsection{Correctness Properties}
\label{sec:correctness-properties}

This section aims to give a complete, but informal, list of high-level correctness properties that a \verb|SimpleMultiSig| implementation should have.

\begin{itemize}
\item If the transaction does not include a call payload, then the transaction is rejected and has no effect on the Ethereum state.
\item The threshold and set of owners can never be changed once the program is deployed to an account.
\item The program is not susceptible to replay attacks. Once a transaction’s call payload is executed successfully on-chain, the transaction, or a part thereof, can never be used to have any further effect on the Ethereum state.
\item If the transaction does not have at least a threshold number of owner signatures that each sign the transaction’s call payload, the transaction has no effect on the Ethereum state.
\item If the transaction does have at least a threshold number of owner signatures that each sign the transaction’s call payload, and the call payload is invoked exactly once intraprocedurally.
\item If the transaction’s call payload is invoked but does not execute successfully, the transaction has no effect on the Ethereum state.
\item The program is effectively callback free~\cite{Grossman:2018} and thus has no re-entrancy vulnerabilities.
\item Ether is never locked in the program.
\item The program has no other functionality.
\end{itemize}

\subsection{Modifications}
I made two changes to the SimpleMultiSig implementation to make it easier to formally verify using KEVM and called this new version \verb|SimpleMultiSigT3| as shown in Listing~\ref{lst:simplemultisigt3} in the appendix.

\subsubsection{Modification \#1: Statically-sized Arrays}

The \verb|SimpleMultiSig| uses dynamically-sized array parameters in the \verb|execute| function to pass the $(V, R, S)$ components of the signatures. In principle, KEVM can be used to reason about dynamic arrays, however, to \emph{automate} this reasoning, KEVM would need to be extended with additional lemmas, which are difficult for a non-expert to write correctly. Thus, I modified the code by replacing these dynamically-sized array parameters with statically-sized arrays instead, which means that the program would only work for a specific threshold. For example, Listing~\ref{lst:simplemultisigt3} in the appendix shows the program would change when the threshold is statically set to 3.

Fortunately, this modification does not restrict the functionality of the \verb|SimpleMultiSig| because the threshold is immutable anyway once the constructor initializes the contract account. However, it does mean that users would need to code-generate the program for the specific threshold that they needed. Incidentally, using statically-sized arrays also has the advantage of reducing gas costs by:

\begin{itemize}
\item reducing transaction size since the ABI encoding of the static arrays is shorter, 
\item reducing overhead of copying the array calldata into memory, 
\item eliminating the two \verb|require| checks on the lengths of the signature parameters since the lengths are statically-enforced,
\item eliminating memory reads on the signature array-length checks in signature validation loop, and 
\item reducing the size of the bytecode
\end{itemize}

To simplify the presentation in this paper, I use the specific code-generation of the SimpleMultiSig, named to \verb|SimpleMultiSigT3|, where the threshold is statically fixed to three, but the discussion and the specification can be straightforwardly adapted to any threshold.

\subsubsection{Modification \#2: Solidity version}

I also replaced the directive \verb|pragma solidity ^0.4.24| with \verb|pragma solidity 0.5.0| to restrict the scope of the verification to EVM bytecode generated by a Solidity 0.5.0 compiler. The Solidity 0.4.24 compiler is missing some code optimizations that the 0.5.0 version introduces. These optimizations generate significantly simpler bytecode which was easier for KEVM to reason about automatically as discussed in Section~\ref{sec:encodepacked-keccak01}.

\subsection{Partial Formal Specification of SimpleMultiSigT3}
\label{sec:specification}
The partial formal specification written for \verb|SimpleMultiSigT3| is given in Listing~\ref{lst:simplemultisigt3-spec} in the Appendix. The specification consists of seven individual K rules informally described below.

\begin{itemize}
\item The \verb|executor-invalid| rule checks that the \verb|execute| function reverts if the \verb|executor| argument is not 0 and is not equal to the \verb|msg.sender| of the transaction.
\item The \verb|sigcheck-fail-revert-0| rule checks that the \verb|execute| function reverts if the first signature is not valid for the EIP712 encoding of the multisig payload, which includes the  \verb|gasLimit|, \verb|destination|, \verb|executor|, \verb|value|, and \verb|data| arguments, and the \verb|nonce| storage variable.
\item The \verb|sigcheck-fail-revert-1| rule is the same as the previous rule but for the second signature.
\item The \verb|sigcheck-fail-revert-2| rule is the same as the previous rule but for the third signature.
\item The \verb|ownercheck-fail-revert| rule checks that the \verb|execute| function reverts if any of the three addresses recovered from the signatures are not owners. And, the function reverts if the three addresses are not in a strictly increasing order. In other words, the first address must be numerically less than the second address, and the second address must be numerically less than the third address. This requirement ensures the addresses are unique.
\item The \verb|call-failure| rule checks that the \verb|execute| function reverts if the call to the \verb|destination| argument returns 0, which indicates the call reverted, for example, due to running out of gas or other reasons. 
\item The \verb|call-success| rule checks that the \verb|execute| function succeeds if 1) the \verb|executor| argument is either 0 or is equal to \verb|msg.sender|, 2) the three signatures are all valid and each recover to unique owners of the wallet, 3) the call invokes the \verb|destination| exactly once and with the correct multisig payload, 4) the call returns 1, which indicates that it succeeded, and 5) no storage variables are modified after the call.
\end{itemize}

Note that these rules do not comprehensively cover all the correctness properties listed in Section~\ref{sec:correctness-properties}. For example, they do not check the following aspects of the SimpleMultiSig, among others, which we leave for future work:
\begin{itemize}
\item storage initialization by constructor
\item ether balance updates
\item re-entrancy
\item bounds on gas usage
\item dynamic array calldata
\end{itemize}
Furthermore, a full formal verification would also require informally, yet rigorously, arguing that the formal specification faithfully captures the correctness properties in Section~\ref{sec:correctness-properties}.

The table below gives the results of running the seven K rules through the prover over the \verb|SimpleMultiSigT3| program using a c5.2xlarge AWS EC2 instance, which has 16GB of RAM and a 3.4 GHz Intel Xeon Platinum 8000 processor.

\begin{table}[!htbp]
\centering
\begin{tabular}{ |l|c|c| } 
\hline
 \textbf{Rule} & \textbf{Time (min)} & \textbf{Result}  \\
 \hline
 \verb|executor-invalid| & 2.2  & \cellcolor[HTML]{D4EDDA} proved true\\ \hline
 \verb|sigcheck-fail-revert-0| & 7.1 & \cellcolor[HTML]{D4EDDA} proved true \\ \hline
 \verb|sigcheck-fail-revert-1| & 11.2 & \cellcolor[HTML]{D4EDDA} proved true \\ \hline
 \verb|sigcheck-fail-revert-2| & 18.9 & \cellcolor[HTML]{D4EDDA} proved true \\ \hline
 \verb|ownercheck-fail-revert| & 44.2 & \cellcolor[HTML]{D4EDDA} proved true \\ \hline
 \verb|call-failure| & 45.9  & \cellcolor[HTML]{D4EDDA} proved true \\ \hline
 \verb|call-success| & 48.6  & \cellcolor[HTML]{D4EDDA} proved true \\ 
 \hline
\end{tabular}
\caption{Results of running the seven K rules through the prover over SimpleMultiSigT3.}
\label{table:tbl-proof-results}
\end{table}

\section{Testing Specifications}
\label{sec:testing}
The output of the prover is a simple "yes" or "no": was it, or was it not, able to prove the specification? This meant that I had to trust that the prover itself did not have bugs, but, as is usually the case with formal verification provers, their implementations are large,  extremely complex, and depend on sophisticated mathematics and algorithms which could easily have been coded incorrectly (Section~\ref{sec:trust}). Because the stakes of a bug in the \verb|SimpleMultiSig| is so high, I had to find a way to at least spot check the prover's work. 

Similarly, I also had to trust that the formal specification I wrote itself did not have any bugs, which is very possible considering how long it is and how unusual specification languages are to a programmer.

To try to mitigate both of these trust issues, I created thirty-two faulty variations of the \verb|SimpleMultiSigT3| program. In the software testing literature, these faulty variations are called \textit{mutants}~\cite{jia:2019} and are constructed by taking the original program and changing it slightly to add a bug. Traditionally, mutants have been used to evaluate the effectiveness of manual test suites, but here I use the same idea to spot check the specification and prover. I ran the prover with the formal specification in Section~\ref{sec:specification} over each of the mutants and then checked to make sure that the prover \emph{failed} to prove the specification on each mutant. The remainder of this section discusses a few examples of the different types of mutants used, and the full list of the test results can be found at \verb|https://www.kspec.io/simplemultisig/tests.html|.

\subsection{Example Test Case: Call Mutation}
The \verb|call_7.sol| mutant adds a bug to \verb|SimpleMultiSigT3| by adding a second call to the \verb|destination|:

\includegraphics[width=.9\linewidth]{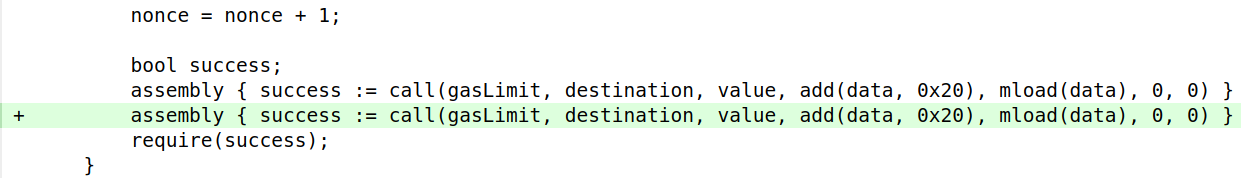}

Table~\ref{tbl:call-7-results} below gives the results of running the prover on this mutant. This test case successfully passes because at least one of the K rules fail, specifically, the \verb|call-failure| and the \verb|call-success| rules.

\begin{table}[!htbp]
\centering
\begin{tabular}{ |l|c|c| } 
\hline
 \textbf{Rule} & \textbf{Time (min)} & \textbf{Result}  \\
 \hline
 \verb|executor-invalid| & 2.0  & \cellcolor[HTML]{D4EDDA} proved true\\ \hline
 \verb|sigcheck-fail-revert-0| & 6.6 & \cellcolor[HTML]{D4EDDA} proved true \\ \hline
 \verb|sigcheck-fail-revert-1| & 11.3 & \cellcolor[HTML]{D4EDDA} proved true \\ \hline
 \verb|sigcheck-fail-revert-2| & 19.2 & \cellcolor[HTML]{D4EDDA} proved true \\ \hline
 \verb|ownercheck-fail-revert| & 42.5 & \cellcolor[HTML]{D4EDDA} proved true \\ \hline
 \verb|call-failure| & 51.6  & \cellcolor[HTML]{F8D7DA} error \\ \hline
 \verb|call-success| & 53.0  & \cellcolor[HTML]{F8D7DA} error \\ 
 \hline
\end{tabular}
\caption{Results of running the seven K rules through the prover over a mutant which invokes the destination address more than once.}
\label{tbl:call-7-results}
\end{table}

\subsection{Example Test Case: EIP712 Encoding Mutation}
The \verb|eip712_0.sol| mutant adds a bug to the EIP712 encoding of the multisig payload by changing the first argument of the \verb|abi.encoding| function invocation to 0 instead of \verb|TXHASH_TYPE|

\includegraphics[width=.9\linewidth]{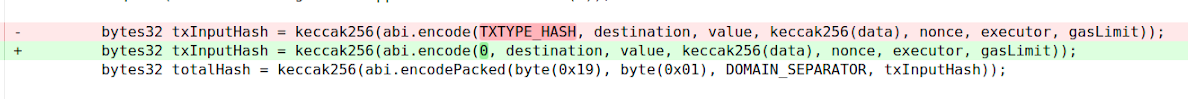}

Table~\ref{tbl:eip712-0-results} below gives the results of running the K prover on this mutant. All except one of the specifications time out, which means that the prover process failed to exit within the three-hour time limit. I chose three hours because, for the correct \verb|SimpleMultiSigT3.sol| program, the prover exits in under an hour for each of the specifications. I make the assumption that if the prover takes at least three times longer than it would on the correct program, then it will not be able to prove the specification on the incorrect program even if it was run for longer than three hours.

\begin{table}[!htbp]
\centering
\begin{tabular}{ |l|c|c| } 
\hline
 \textbf{Rule} & \textbf{Time (min)} & \textbf{Result}  \\
 \hline
 \verb|executor-invalid| & 2.2  & \cellcolor[HTML]{D4EDDA} proved true\\ \hline
 \verb|sigcheck-fail-revert-0| & 180 & \cellcolor[HTML]{F8D7DA} timeout \\ \hline
 \verb|sigcheck-fail-revert-1| & 180 & \cellcolor[HTML]{F8D7DA} timeout \\ \hline
 \verb|sigcheck-fail-revert-2| & 180 & \cellcolor[HTML]{F8D7DA} timeout \\ \hline
 \verb|ownercheck-fail-revert| & 180 & \cellcolor[HTML]{F8D7DA} timeout \\ \hline
 \verb|call-failure| & 180  & \cellcolor[HTML]{F8D7DA} timeout \\ \hline
 \verb|call-success| & 180  & \cellcolor[HTML]{F8D7DA} timeout \\ 
 \hline
\end{tabular}
\caption{Results of running the seven K rules through the prover on a mutant which incorrectly passes the value 0 to the EIP712 encoding rather than TXHASH\_TYPE.}
\label{tbl:eip712-0-results}
\end{table}

\subsection{Example Test Case: Signature Checking Mutation}
The \verb|sigcheck_5.sol| mutant adds a bug to \verb|SimpleMultiSigT3| by removing the \verb|requires| check which ensures the signatures are unique by enforcing that they are passed to the function in order of the strictly increasing value of their recovered addresses:

\includegraphics[width=.9\linewidth]{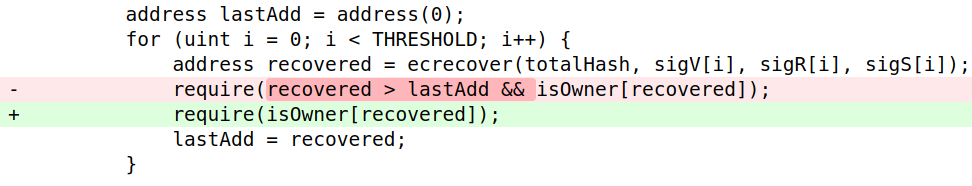}

Table~\ref{tbl:sigcheck-5-results} below gives the results of running the prover on this mutant. This test case successfully passes because at least one of the K rules fail, specifically, all of them except for the \verb|executor-invalid| spec.

\begin{table}[!htbp]
\centering
\begin{tabular}{ |l|c|c| } 
\hline
 \textbf{Rule} & \textbf{Time (min)} & \textbf{Result}  \\
 \hline
 \verb|executor-invalid| & 2.1  & \cellcolor[HTML]{D4EDDA} proved true\\ \hline
 \verb|sigcheck-fail-revert-0| & 180.0 & \cellcolor[HTML]{F8D7DA} timeout \\ \hline
 \verb|sigcheck-fail-revert-1| & 30.9 & \cellcolor[HTML]{F8D7DA} error \\ \hline
 \verb|sigcheck-fail-revert-2| & 24.1 & \cellcolor[HTML]{F8D7DA} error \\ \hline
 \verb|ownercheck-fail-revert| & 49.6 & \cellcolor[HTML]{F8D7DA} error \\ \hline
 \verb|call-failure| & 46.9  & \cellcolor[HTML]{F8D7DA} error \\ \hline
 \verb|call-success| & 50.0  & \cellcolor[HTML]{F8D7DA} error \\ 
 \hline
\end{tabular}
\caption{Results of running the seven K rules through the prover over a mutant which removes the uniqueness check on the signatures.}
\label{tbl:sigcheck-5-results}
\end{table}

\begin{acks}
This work was funded by ConsenSys R\&D. Thanks to Mario Alvarez, Joseph Chow, Robert Drost, Christian Lundkvist, and Valentin Wüstholz for their thoughtful feedback throughout the project. Thanks to Grigore Rosu for his support of this work, his commitment to open research and generously open-sourcing and liberally licensing the K Framework. Thanks to Denis Bogdanas, Dwight Guth, Everett Hildenbrandt, Daejun Park, and Yi Zhang for answering my technical questions about the K prover and devising the additional lemmas needed to push some of the proofs through.
\end{acks}

\bibliographystyle{ACM-Reference-Format}
\bibliography{paper}

\appendix

\begin{lstlisting}[label={lst:simplemultisig},language=Solidity, caption=Original implementation of the SimpleMultiSig smart contract~\cite{Lundqvist:2015},float=*]
pragma solidity ^0.4.24;

contract SimpleMultiSig {

  // EIP712 Precomputed hashes:
  // keccak256("EIP712Domain(string name,string version,uint256 chainId,address verifyingContract,bytes32 salt)")
  bytes32 constant EIP712DOMAINTYPE_HASH = 0xd87cd6ef79d4e2b95e15ce8abf732db51ec771f1ca2edccf22a46c729ac56472;

  // keccak256("Simple MultiSig")
  bytes32 constant NAME_HASH = 0xb7a0bfa1b79f2443f4d73ebb9259cddbcd510b18be6fc4da7d1aa7b1786e73e6;

  // keccak256("1")
  bytes32 constant VERSION_HASH = 0xc89efdaa54c0f20c7adf612882df0950f5a951637e0307cdcb4c672f298b8bc6;

  // keccak256("MultiSigTransaction(address destination,uint256 value,bytes data,uint256 nonce,address executor,uint256 gasLimit)")
  bytes32 constant TXTYPE_HASH = 0x3ee892349ae4bbe61dce18f95115b5dc02daf49204cc602458cd4c1f540d56d7;

  bytes32 constant SALT = 0x251543af6a222378665a76fe38dbceae4871a070b7fdaf5c6c30cf758dc33cc0;

  uint public nonce;                 // (only) mutable state
  uint public threshold;             // immutable state
  mapping (address => bool) isOwner; // immutable state
  address[] public ownersArr;        // immutable state

  bytes32 DOMAIN_SEPARATOR;          // hash for EIP712, computed from contract address
  
  // Note that owners_ must be strictly increasing, in order to prevent duplicates
  constructor(uint threshold_, address[] owners_, uint chainId) public {
    require(owners_.length <= 10 && threshold_ <= owners_.length && threshold_ > 0);

    address lastAdd = address(0);
    for (uint i = 0; i < owners_.length; i++) {
      require(owners_[i] > lastAdd);
      isOwner[owners_[i]] = true;
      lastAdd = owners_[i];
    }
    ownersArr = owners_;
    threshold = threshold_;

    DOMAIN_SEPARATOR = keccak256(abi.encode(EIP712DOMAINTYPE_HASH,NAME_HASH,VERSION_HASH,chainId,this,SALT));
  }

  // Note that address recovered from signatures must be strictly increasing, in order to prevent duplicates
  function execute(uint8[] sigV, bytes32[] sigR, bytes32[] sigS, address destination, uint value, bytes data, address executor, uint gasLimit) public {
    require(sigR.length == threshold);
    require(sigR.length == sigS.length && sigR.length == sigV.length);
    require(executor == msg.sender || executor == address(0));

    // EIP712 scheme: https://github.com/ethereum/EIPs/blob/master/EIPS/eip-712.md
    bytes32 txInputHash = keccak256(abi.encode(TXTYPE_HASH, destination, value, keccak256(data), nonce, executor, gasLimit));
    bytes32 totalHash = keccak256(abi.encodePacked("\x19\x01", DOMAIN_SEPARATOR, txInputHash));

    address lastAdd = address(0); // cannot have address(0) as an owner
    for (uint i = 0; i < threshold; i++) {
      address recovered = ecrecover(totalHash, sigV[i], sigR[i], sigS[i]);
      require(recovered > lastAdd && isOwner[recovered]);
      lastAdd = recovered;
    }

    // If we make it here all signatures are accounted for.
    nonce = nonce + 1;
    bool success = false;
    assembly { success := call(gasLimit, destination, value, add(data, 0x20), mload(data), 0, 0) }
    require(success);
  }

  function () payable external {}
}
\end{lstlisting}

\begin{lstlisting}[label={lst:simplemultisigt3}, language=Solidity, caption=SimpleMultiSigT3,float=*]
pragma solidity 0.5.0;

contract SimpleMultiSigT3 {

    // EIP712 Precomputed hashes:
    // keccak256("EIP712Domain(string name,string version,uint256 chainId,address verifyingContract,bytes32 salt)")
    bytes32 constant EIP712DOMAINTYPE_HASH = 0xd87cd6ef79d4e2b95e15ce8abf732db51ec771f1ca2edccf22a46c729ac56472;

    // keccak256("Simple MultiSig")
    bytes32 constant NAME_HASH = 0xb7a0bfa1b79f2443f4d73ebb9259cddbcd510b18be6fc4da7d1aa7b1786e73e6;

    // keccak256("1")
    bytes32 constant VERSION_HASH = 0xc89efdaa54c0f20c7adf612882df0950f5a951637e0307cdcb4c672f298b8bc6;

    // keccak256("MultiSigTransaction(address destination,uint256 value,bytes data,uint256 nonce,address executor,uint256 gasLimit)")
    bytes32 constant TXTYPE_HASH = 0x3ee892349ae4bbe61dce18f95115b5dc02daf49204cc602458cd4c1f540d56d7;

    bytes32 constant SALT = 0x251543af6a222378665a76fe38dbceae4871a070b7fdaf5c6c30cf758dc33cc0;

    uint constant THRESHOLD = 3;       // immutable state

    uint public nonce;                 // (only) mutable state
    mapping (address => bool) isOwner; // immutable state
    address[] public ownersArr;        // immutable state
    bytes32 DOMAIN_SEPARATOR;          // hash for EIP712, computed from contract address

    // Note that owners_ must be strictly increasing, in order to prevent duplicates
    constructor(address[] memory owners_, uint chainId) public {
        require(owners_.length <= 10 && THRESHOLD <= owners_.length);

        address lastAdd = address(0);
        for (uint i = 0; i < owners_.length; i++) {
            require(owners_[i] > lastAdd);
            isOwner[owners_[i]] = true;
            lastAdd = owners_[i];
        }
        ownersArr = owners_;

        DOMAIN_SEPARATOR = keccak256(abi.encode(EIP712DOMAINTYPE_HASH,NAME_HASH,VERSION_HASH,chainId,this,SALT));
    }

    // Note that address recovered from signatures must be strictly increasing, in order to prevent duplicates
    function execute(uint8[3] memory sigV, bytes32[3] memory sigR, bytes32[3] memory sigS, address destination, uint value, bytes memory data, address executor, uint gasLimit) public {
        require(executor == msg.sender || executor == address(0));

        // EIP712 scheme: https://github.com/ethereum/EIPs/blob/master/EIPS/eip-712.md
        bytes32 txInputHash = keccak256(abi.encode(TXTYPE_HASH, destination, value, keccak256(data), nonce, executor, gasLimit));
        bytes32 totalHash = keccak256(abi.encodePacked("\x19\x01", DOMAIN_SEPARATOR, txInputHash));

        address lastAdd = address(0); // cannot have address(0) as an owner
        for (uint i = 0; i < THRESHOLD; i++) {
            address recovered = ecrecover(totalHash, sigV[i], sigR[i], sigS[i]);
            require(recovered > lastAdd && isOwner[recovered]);
            lastAdd = recovered;
        }

        // If we make it here all signatures are accounted for.
        nonce = nonce + 1;
        bool success = false;
        assembly { success := call(gasLimit, destination, value, add(data, 0x20), mload(data), 0, 0) }
        require(success);
    }

    function () payable external {}
}
\end{lstlisting}

\begin{lstlisting}[firstline=1, lastline=75, label={lst:simplemultisigt3-spec}, language=KYaml, caption=Partial formal specification of SimpleMultiSigT3, float=*]
spec:
  - rule:
      name: base
      if:
        match:
          storage: S0
          callData:  \#abiCallData2("execute(uint8[3],bytes32[3],bytes32[3],address,uint256,bytes,address,uint256)",
            \#uint8(SIGV0), \#uint8(SIGV1), \#uint8(SIGV2),
            \#bytes32(SIGR0), \#bytes32(SIGR1), \#bytes32(SIGR2),
            \#bytes32(SIGS0), \#bytes32(SIGS1), \#bytes32(SIGS2),
            \#address(DESTINATION),
            \#uint256(VALUE),
            \#bytes(\#buf(DATA_LEN,DATA)),
            \#address(EXECUTOR),
            \#uint256(GAS_LIMIT))
          call_log_pcset: SetItem(CALL_PC) .Set
        where:
          - CALL_PC ==Int 1786
          - TXTYPE_HASH ==K "0x3ee892349ae4bbe61dce18f95115b5dc02daf49204cc602458cd4c1f540d56d7"
          - NONCE_SLOT ==Int 0
          - NONCE_0 ==Int select(S0, \#hashedLocation("Solidity", NONCE_SLOT, .IntList))
          - \#rangeUInt(256, NONCE_0)
          - NONCE_0 <Int 2 ^Int 256 -Int 1
          - OWNER_MAPPING_SLOT ==Int 1
          - DOMAIN_SEPARATOR ==Int select(S0, \#hashedLocation("Solidity", 3, .IntList))
          - \#rangeAddress(DESTINATION)
          - \#rangeUInt(256, VALUE)
          - \#rangeUInt(256, DATA_LEN)
          - DATA_LEN <Int 2 ^Int 16
          - \#rangeAddress(EXECUTOR)
          - \#rangeUInt(256, GAS_LIMIT)
          - \#rangeBytes(32, DOMAIN_SEPARATOR)
          - \#rangeUInt(8, SIGV0)
          - \#rangeUInt(8, SIGV1)
          - \#rangeUInt(8, SIGV2)
          - \#rangeBytes(32, SIGR0)
          - \#rangeBytes(32, SIGR1)
          - \#rangeBytes(32, SIGR2)
          - \#rangeBytes(32, SIGS0)
          - \#rangeBytes(32, SIGS1)
          - \#rangeBytes(32, SIGS2)

  - rule:
      name: executor-invalid
      inherits: base
      if:
        where:
          - EXECUTOR =/=Int MSG_SENDER
          - EXECUTOR =/=Int 0
      then:
        match:
          statusCode: EVMC_REVERT

  - rule:
      name: executor-valid
      inherits: base
      if:
        where:
          - or:
              - EXECUTOR ==Int MSG_SENDER
              - EXECUTOR ==Int 0
          - TX_INPUT_HASH ==Int keccak(\#encodeArgs(
            \#bytes32(\#parseHexWord(TXTYPE_HASH)),
            \#address(DESTINATION),
            \#uint256(VALUE),
            \#bytes32(keccak(\#buf(DATA_LEN,DATA))),
            \#uint256(NONCE_0),
            \#address(EXECUTOR),
            \#uint256(GAS_LIMIT)))
          - >
            TX_TOTAL_HASH ==Int keccak(25: 1: #encodeArgs(#bytes32(DOMAIN_SEPARATOR),#bytes32(TX_INPUT_HASH)))
          - ECREC_DATA0 ==K \#encodeArgs( \#bytes32(TX_TOTAL_HASH), \#uint8(SIGV0), \#bytes32(SIGR0), \#bytes32(SIGS0))
          - ECREC_DATA1 ==K \#encodeArgs( \#bytes32(TX_TOTAL_HASH), \#uint8(SIGV1), \#bytes32(SIGR1), \#bytes32(SIGS1))
          - ECREC_DATA2 ==K \#encodeArgs( \#bytes32(TX_TOTAL_HASH), \#uint8(SIGV2), \#bytes32(SIGR2), \#bytes32(SIGS2))
\end{lstlisting}

\begin{lstlisting}[firstnumber=76, firstline=76, lastline=144, language=KYaml, caption=Partial formal specification of SimpleMultiSigT3, float=*]
  - rule:
      name: sigcheck-fail-revert-0
      inherits: executor-valid
      if:
        where:
          - \#ecrecEmpty( ECREC_DATA0 )
      then:
        match:
          statusCode: EVMC_REVERT

  - rule:
      name: valid-sigcheck-fail-revert-1
      inherits: executor-valid
      if:
        where:
          - not: \#ecrecEmpty( ECREC_DATA0 )
          - \#ecrecEmpty( ECREC_DATA1 )
      then:
        match:
          statusCode: EVMC_REVERT

  - rule:
      name: valid-sigcheck-fail-revert-2
      inherits: executor-valid
      if:
        where:
          - not: \#ecrecEmpty( ECREC_DATA0 )
          - not: \#ecrecEmpty( ECREC_DATA1 )
          - \#ecrecEmpty( ECREC_DATA2 )
      then:
        match:
          statusCode: EVMC_REVERT

  - rule:
      name: sigcheck-pass
      inherits: executor-valid
      if:
        where:
          - RECOVERED0 ==Int \#symEcrec( ECREC_DATA0 )
          - RECOVERED1 ==Int \#symEcrec( ECREC_DATA1 )
          - RECOVERED2 ==Int \#symEcrec( ECREC_DATA2 )
          - not: \#ecrecEmpty( ECREC_DATA0 )
          - not: \#ecrecEmpty( ECREC_DATA1 )
          - not: \#ecrecEmpty( ECREC_DATA2 )
          - \#rangeAddress(RECOVERED0)
          - \#rangeAddress(RECOVERED1)
          - \#rangeAddress(RECOVERED2)
          - IS_OWNER0 ==Int select(S0, \#hashedLocation("Solidity", OWNER_MAPPING_SLOT, RECOVERED0))
          - IS_OWNER1 ==Int select(S0, \#hashedLocation("Solidity", OWNER_MAPPING_SLOT, RECOVERED1))
          - IS_OWNER2 ==Int select(S0, \#hashedLocation("Solidity", OWNER_MAPPING_SLOT, RECOVERED2))
          - \#rangeUInt(8, IS_OWNER0)
          - \#rangeUInt(8, IS_OWNER1)
          - \#rangeUInt(8, IS_OWNER2)

  - rule:
      name: ownercheck-fail-revert
      inherits: sigcheck-pass
      if:
        where:
          - or:
              - IS_OWNER0 ==Int 0
              - IS_OWNER1 ==Int 0
              - IS_OWNER2 ==Int 0
              - RECOVERED0 >=Int RECOVERED1
              - RECOVERED1 >=Int RECOVERED2
      then:
        match:
          statusCode: EVMC_REVERT
\end{lstlisting}

\begin{lstlisting}[firstnumber=145, firstline=145, lastline=225, language=KYaml, caption=Partial formal specification of SimpleMultiSigT3, float=*]
  - rule:
      name: ownercheck-pass-call
      inherits: sigcheck-pass
      if:
        match:
          storage: S0
          readLog: .List
          writeLog: .List
          callLog: .List
        where:
          - IS_OWNER0 =/=Int 0
          - IS_OWNER1 =/=Int 0
          - IS_OWNER2 =/=Int 0
          - RECOVERED0 <Int RECOVERED1
          - RECOVERED1 <Int RECOVERED2
      then:
        match:
          storage: S1
          readLog: RD_FULL
          writeLog: WR_FULL
          callLog: ListItem ( 0 CALL_PC GAS_LIMIT DESTINATION VALUE ARGSTART_C ARGWIDTH_C 0 0 LM_C RD_C WR_C) .List
        where:
          - selectRange(LM_C, ARGSTART_C, ARGWIDTH_C) ==K \#buf(DATA_LEN,DATA)
          - S0 ==IMap S1 except (SetItem(\#hashedLocation("Solidity", NONCE_SLOT, .IntList)) .Set)
          - NONCE_1 ==Int select(S1, \#hashedLocation("Solidity", NONCE_SLOT, .IntList))
          - NONCE_1 >Int NONCE_0
          - RD_FULL ==K RD_C
          - WR_FULL ==K WR_C

  - rule:
      name: call-success
      inherits: ownercheck-pass-call
      if:
        where:
          - \#callSuccess(0, DESTINATION)
      then:
        match:
          statusCode: EVMC_SUCCESS

  - rule:
      name: call-failure
      inherits: ownercheck-pass-call
      if:
        where:
          - \#callFailure(0, DESTINATION)
      then:
        match:
          statusCode: EVMC_REVERT
\end{lstlisting}

\end{document}